\documentclass[12pt,headings=big,numbers=noenddot,DIV=14,a4paper]{scrartcl}%

\pdfoutput=1

\usepackage{bbm}
\usepackage{amsmath,amssymb,amsfonts}
\usepackage[normal,font=small,labelfont=bf,labelsep=period]{caption}
\usepackage[pdftex]{color,graphicx}
\usepackage{subfigure}
\usepackage[english]{babel}
\usepackage[compress]{cite}
\addtokomafont{disposition}{\rmfamily\boldmath}

\usepackage[dvipsnames]{xcolor}
\definecolor{darkblue}{rgb}{0,0.2,0.6}
\definecolor{darkgreen}{rgb}{0,0.4,0}

\usepackage[linktoc=page,bookmarks=false,colorlinks=false,
linkbordercolor=RoyalBlue,citebordercolor=ForestGreen,urlbordercolor=CornflowerBlue]{hyperref}

\numberwithin{equation}{section}
\newcommand{\gsim}{\lower.7ex\hbox{$\;\stackrel{\textstyle>}{\sim}\;$}}
\newcommand{\lsim}{\lower.7ex\hbox{$\;\stackrel{\textstyle<}{\sim}\;$}}

\DeclareFontFamily{OT1}{pzc}{}
\DeclareFontShape{OT1}{pzc}{m}{it}{<-> s * [1.10] pzcmi7t}{}
\DeclareMathAlphabet{\mathpzc}{OT1}{pzc}{m}{it}

\title{\vskip 30pt
       \textcolor{black}{The Composite Twin Higgs scenario}}
\vspace{0.4cm}
\author{{\small{Riccardo Barbieri$^1$\footnote{riccardo.barbieri@sns.it} , Davide Greco$^2$\footnote{davide.greco@epfl.ch} , Riccardo Rattazzi$^2$\footnote{riccardo.rattazzi@epfl.ch}  and Andrea Wulzer$^3$\footnote{andrea.wulzer@pd.infn.it}}} \\ 
{\small\emph{$^1$Scuola Normale Superiore and INFN, Piazza dei Cavalieri 7, 56126 Pisa, Italy}}\\ 
{\small\emph{$^2$Institut de Th\'{e}orie des Ph\'{e}nom\`{e}nes Physiques,
        EPFL, Lausanne, Switzerland}}\\ 
{\small\emph{$^3$Dipartimento di Fisica e Astronomia, Universit\`a di Padova and}}\\
{\small\emph{INFN, Sezione di Padova, via Marzolo 8, I-35131 Padova, Italy
}}}

\date{}

\def\Id{\mathbbm{1}}
\def\beq{\begin{equation}}
\def\eeq{\end{equation}}

\begin{document}
\begin{titlepage}

\maketitle
\thispagestyle{empty}

\begin{abstract}
\centerline{\bf Abstract}\medskip
\noindent
{Based on an explicit model, we propose and discuss the generic features of a possible implementation of the Twin Higgs program in the context of composite Higgs models. We find that the Twin Higgs quadratic divergence cancellation argument can be uplifted to a genuine protection of the Higgs potential, based on symmetries and selection rules, but only under certain conditions which are not fulfilled in some of the existing models. We also find that a viable scenario, not plagued by a massless Twin Photon, can be obtained by not gauging the Twin Hypercharge and taking this as the only source of Twin Symmetry breaking at a very high scale.}
\end{abstract}

\end{titlepage}

\section{Introduction}

The possibility that there exist models of electroweak symmetry breaking with a minimal amount of fine tuning (less than $10\%$ or so) and the simultaneous absence below a few TeV  of any new particle charged under the Standard Model (SM) gauge group deserves attention. 
Generically the idea behind this possibility goes under the name of Twin Higgs. In this note we discuss an explicit example where this idea is implemented in the context of a composite Higgs picture. We do that with the purpose of proposing and analyzing a few generic features of such an implementation, which will be illustrated in the course of the exposition. 

\section{A model example}

\begin{figure}
\centering
\includegraphics[width=.5\textwidth]{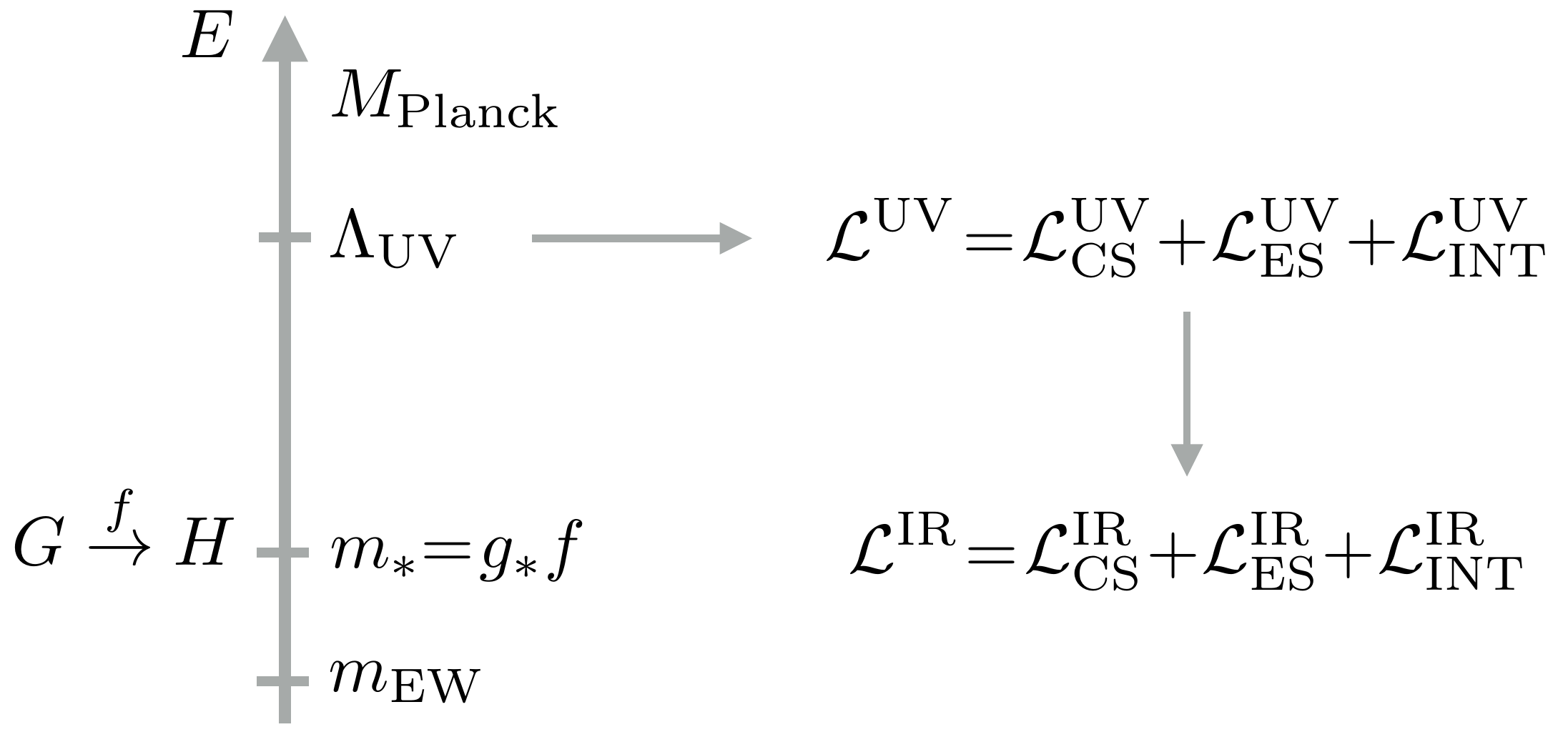}
\caption{A pictorial view of the Composite Higgs framework.}
\label{scaleferm}
\end{figure}

The situation which we have in mind, depicted in Figure~\ref{scaleferm}, is that there exist a new ``Composite Sector" (CS), endowed with a global symmetry group $G$, which confines at a scale $m_*$ in the TeV or multi--TeV range. In the process, $G$ gets spontaneously broken to a subgroup $H$ and the order parameter for this breaking, $f$, is related to the confinement scale by $m_*=g_*f$. The scale $m_*$ sets the typical mass of the Composite Sector resonances and $g_*$ sets their typical interaction strength \cite{Giudice:2007fh}. The Composite Sector itself originates from some unspecified dynamics at a very high scale ${\Lambda_{\textrm{UV}}}\gg m_*$ and the large separation among these two scales is ensured by the hypothesis that the Composite Sector flows toward a conformal fixed point below ${\Lambda_{\textrm{UV}}}$ and it remains close to it until $m_*$. Also one ``Elementary Sector" (ES) is generated at the high scale ${\Lambda_{\textrm{UV}}}$. The latter is composed of weakly--interacting fields, among which the SM ones with the possible exception of the right--handed Top quark, which could also be a fully composite degree of freedom originating from the CS. In the ordinary, or Minimal \cite{Agashe:2004rs}, Composite Higgs construction, the ES   comprises just the SM fields. Instead, as described below, in the Twin Composite Higgs,  the ES also comprises Extra ``Twin" degrees of freedom. The CS does exactly respect  $G$ invariance, but the ES breaks it badly because its degrees of freedom do not come in $G$ multiplets. Explicit $G$ symmetry--breaking effects are communicated to the CS through the Elementary/Composite interactions, denoted as ${\mathcal{L}}_{\text{INT}}$ in Figure~\ref{scaleferm}. They come as weak interactions at ${\Lambda_{\textrm{UV}}}$ and they are assumed not to be strongly relevant operators such as to remain weak when evolved down at the IR scale $m_*$. Therefore it makes sense to treat perturbatively their effects on the IR dynamics as tiny $G$--breaking perturbations. 

Let us now come to our specific construction. The relevant global symmetry group of the CS is $SO(8)$, which gets spontaneously broken to an $SO(7)$ subgroup delivering $7$ Goldstone Bosons in the $\mathbf{7}$ of the unbroken $SO(7)$, out of which only the Higgs boson will survive as a physical particle. A total of $7$ Elementary gauge fields are introduced, and coupled to the CS by weakly gauging $7$ of the $28$ $SO(8)$ generators, whose explicit form is reported in Appendix~\ref{SO(8)Gen} for the Fundamental representation. In particular, we gauge some of the generators which live in the block--diagonal $SO(4)\times \widetilde{SO}(4)$ subgroup, namely those of the $SU(2)_L\times U(1)_{3,R}$ and $\widetilde{SU}(2)_L$ subgroups of the two $SO(4)\simeq SU(2)_L\times SU(2)_R$. The group $SO(4)$ is taken to be part of the unbroken $SO(7)$, while $\widetilde{SO}(4)$ is partially broken by the CS, namely $\widetilde{SO}(4)\rightarrow \widetilde{SO}(3)$ at the scale $f$. The SM group being embedded in the unbroken $SO(4)$ ensures Custodial protection and avoids unacceptably large tree--level corrections to the ${T}$ parameter of ElectroWeak Precision Tests (EWPT). This Custodial protection is one reason for having an $SO(8)/SO(7)$ spontaneous symmetry breaking pattern in the CS, as already noted in \cite{Geller:2014kta}.

The $SU(2)_L\times U(1)_{3,R}$ group is identified with the electroweak SM gauge group and the corresponding gauge fields thus deliver the EW bosons and the photon. The remaining $3$ elementary vector fields gauging $\widetilde{SU}(2)_L$ correspond instead to new particles, which we call the ``Twin partners'' of the SM $W$ fields. They are associated with generators that  commute with the SM group and are thus  EW--neutral objects. Given that $\widetilde{SU}(2)_L$ is broken by the CS, the Twin $W$'s are massive and acquire their longitudinal components from $3$ of the $7$ Goldstones, which thus disappear from the spectrum. The remaining $4$, associated with the generators $T^{\mathbf{7}}_{1,\ldots,4}$  in Appendix~\ref{SO(8)Gen}, are in the ${\mathbf{4}}$ of $SO(4)$ and they have precisely the SM quantum numbers of the ordinary Higgs doublet. The latter will eventually acquire a vacuum expectation value (VEV), which we take along  $T^{\mathbf{7}}_{4}$, give a mass to the EW bosons and deliver just one physical scalar, the SM Higgs boson. Unlike in the original Twin Higgs proposal \cite{Chacko:2005pe} and in the subsequent literature, \cite{Geller:2014kta, Barbieri:2005ri, Chacko:2005vw, Chacko:2005un, Chang:2006ra, Batra:2008jy, Craig:2013fga, Craig:2014aea, Craig:2014roa, Burdman:2014zta}, no mirror partner is introduced for the SM Hypercharge field in order to avoid the appearance  of an exactly massless Twin photon in the spectrum. 

\subsection{The gauge sector}

Aside from the Higgs, the EW bosons and the Twin $W$'s, extra massive resonances are present, originating as bound states of the CS. They could come in a variety of spin and $SO(7)$ quantum numbers but in particular we do expect some of them to be spin--one vectors and to have the quantum numbers of the global currents associated to the unbroken group $SO(7)$, {\it{i.e.}} to live in the Adjoint. The QCD analog of these particles are the $\rho$ mesons, which are the lightest spin--one hadrons. Vectors in the Adjoint would also appear in a 5d holographic implementation of our setup. It is thus reasonable to take them as representatives of the CS particle content. Therefore, we introduce an Adjoint (the ${\mathbf{21}}$ of $SO(7)$) of vectors $\rho_a$ and we define a $2$--site model, constructed by the standard rules of Ref.~\cite{Panico:2011pw}, to describe their dynamics. We regard this model as a simple illustrative implementation of the Composite Twin Higgs idea. Its Lagrangian reads
\begin{equation}
\label{gl}
\mathcal{L}^{\textrm{gauge}}=- \frac1{4 g_\rho^2} \sum_{a=1}^{21}\rho_{\mu\nu}^a \rho^{\mu\nu}_a+{f^2 \over 4 }\text{Tr}[(D_\mu \Sigma)^t D^\mu \Sigma]  - \frac1{4 g_{2}^2} W_{\mu\nu}^\alpha W^{\mu\nu}_\alpha - \frac1{4 g_{1}^2} B_{\mu\nu} B^{\mu\nu} - \frac1{4 \widetilde{g}_{2}^2} \widetilde{W}_{\mu\nu}^\alpha \widetilde{W}^{\mu\nu}_\alpha\,,
\end{equation}
where $\rho_{\mu\nu}^a$ are the field--strength tensors of the resonance fields --which are treated in the $2$--site model as gauge fields of a local $SO(7)$ group--, $W_{\mu\nu}^\alpha$ and $B_{\mu\nu}$ are the usual SM field--strengths and $\widetilde{W}_{\mu\nu}^\alpha$ those of the $3$ Twin $W$ partners. The field $\Sigma$ is a generic $SO(8)$ matrix containing $28$ real scalar fields. However, $21$ of these can be eliminated by gauge--fixing the local $SO(7)$ associated with the $\rho$'s, making $\Sigma$ become the exponential of the $7$ broken generators only. In this gauge, $\Sigma$ can be interpreted as the Goldstone Matrix of the $SO(8)/SO(7)$ coset, namely
\beq
\displaystyle
\Sigma = U =e^{-{2 i\over f} \Pi^{\alpha} T^{\mathbf{7}}_{\alpha}}\,.
\eeq
All the $7$ remaining scalars, but one, can be eliminated by gauge--fixing the local $\widetilde{SU}(2)_L$ associated with the Twin $W$'s and the broken SM generators. This defines the Unitary Gauge, in which $\Sigma$ reads 
\begin{equation}
\label{golmat}
\Sigma =U=e^{-{2 i\over f}H T^{\bf 7}_4}=\left(
\begin{array}{cccc}
\Id_3 & 0 & 0 & 0 \\
0 &  \cos \frac{H}{f}  & 0 & \sin \frac{H}{f}  \\
 0 & 0 & \Id_3 &  0 \\
 0 & -\sin \frac{H}{f}  &  0 & \cos \frac{H}{f}  \\
\end{array}
\right)\,,
\end{equation}
where $H$ is the real neutral component of the Higgs doublet (times $\sqrt{2}$) which, after EWSB, decomposes in VEV plus physical Higgs fluctuation as $H(x)=V+h(x)$.

It is important to interpret properly the various terms that appear in Eq.~(\ref{gl}). The first one comes purely from the CS and describes the kinetic term  of the resonances and their self--interactions. The corresponding coupling $g_\rho$ is therefore of the order of the typical CS coupling $g_*$. The last three terms are purely Elementary. In accordance with the hypothesis that the ES is weakly--coupled and gives a subdominant correction to the CS dynamics,  the associated couplings are assumed to satisfy  \beq
\label{weC}
g_{1,2 }\sim \widetilde{g}_{2 }\ll g_\rho\sim g_*\,,
\eeq
 The second term is instead a mixed one. It contains both purely CS operators, among which the Goldstone bosons kinetic term and a mass for the $\rho$'s, and Elementary/Composite interactions. Indeed, the covariant derivative of $\Sigma$ reads
\begin{equation}
D_\mu\Sigma =  \partial_\mu \Sigma -i   A_\mu^A T^A \Sigma + i   \Sigma  \rho_{\mu}^a T_{\bf 21}^a\,,
\end{equation}
where we collected in $A_\mu^A$, $A=1,\ldots, 7$, all the Elementary gauge fields appropriately embedded in the Adjoint of $SO(8)$, namely
\beq
 A_\mu^A T^A = W_\mu^\alpha (T_{\bf L})^\alpha + B_\mu (T_{ \bf R})^3 + \widetilde{W}_\mu^\alpha (\widetilde{T}_{\bf L})^\alpha \,,
\eeq
in terms of the generators defined in Appendix \ref{SO(8)Gen}. 

\begin{figure}
\centering
\includegraphics[width=.7\textwidth]{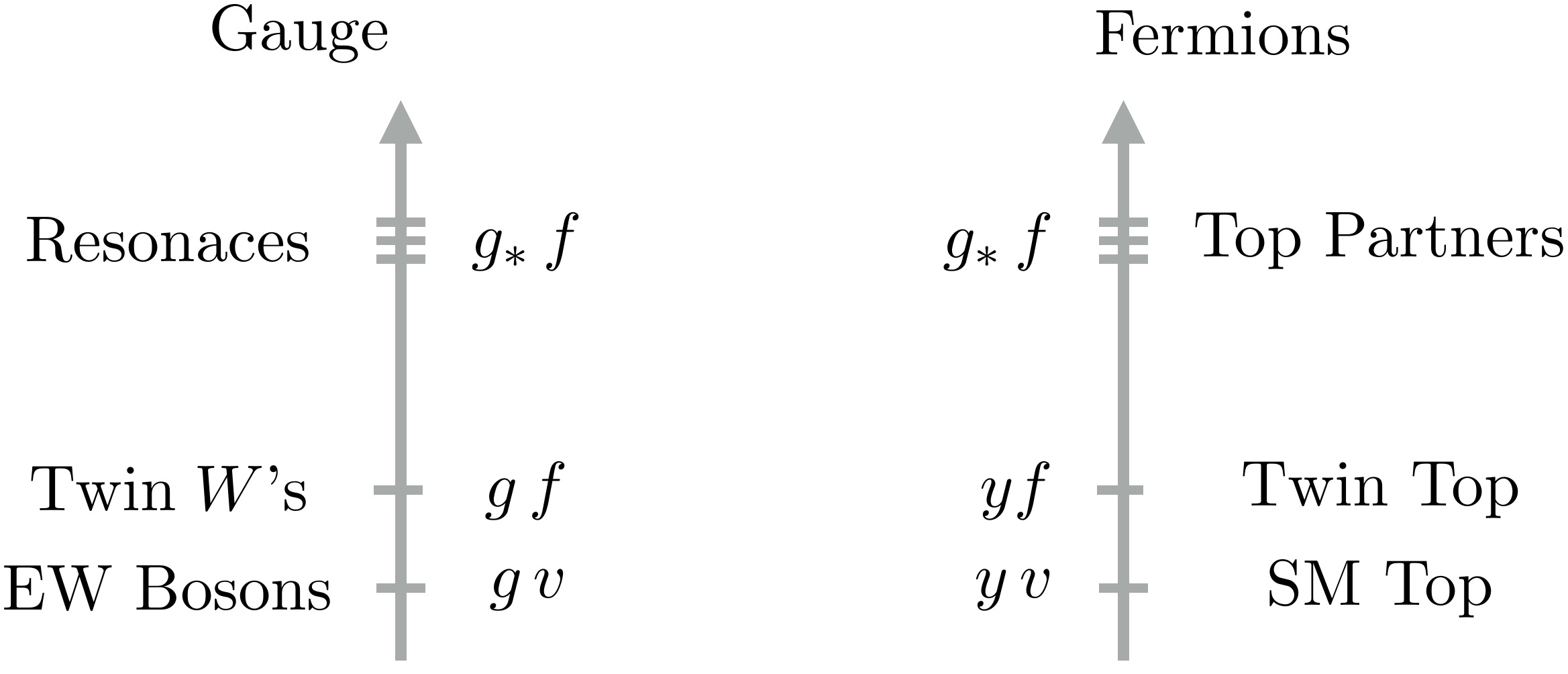}
\caption{The mass spectrum in the gauge (left) and fermionic (right) sectors.}
\label{spec}
\end{figure}

The mass--spectrum of the theory is immediately worked out in the weak Elementary coupling expansion of Eq.~(\ref{weC}). First, we do find the massless photon and the $W$ and $Z$ bosons with masses
\begin{equation}
M_W^2 \simeq \frac{1}{4} g_{2}^2 f^2 \sin^2\frac{V}{f}=\frac14 g_{2}^2v^2\,,\;\;\;\;\;M_Z^2 \simeq \frac{1}{4}  (g_{2}^2+ g_{1}^2) f^2 \sin^2\frac{V}{f}=M_W^2/\cos^2{\theta_W}\,,
\end{equation}
where we identified $g_{1,2}$ with the SM $g_{1,2}$ couplings --which holds up to $g_{1,2}^2/g_\rho^2$ corrections-- and we defined the EWSB scale as
\beq
v=f \sin\frac{V}{f}\simeq246\;{\textrm{GeV}}\,,\;\textrm{thus}\;\;\xi\equiv \frac{v^2}{f^2}=\sin^2\frac{V}{f}\,.
\eeq
Like in the ordinary Composite Higgs setup, we do have plenty of phenomenological reasons to take $\xi$ small. Indeed $\xi$ controls the departures of the Higgs couplings from the SM expectations, which are constrained both from the direct LHC measurements and from their indirect effects on EWPT \cite{Barbieri:2007bh}. The maximal defendable value of $\xi$ is around $0.2$, given that making it small requires fine--tuning in the potential we will take it close to the maximum, which corresponds to a Goldstone scale $f\sim500$~GeV.\footnote{A quantitative compatibility with EWPT is actually possible in ordinary Composite Higgs models only relying on the radiative effects of somewhat light colored Top Partners \cite{Barbieri:2007bh,Grojean:2013qca}, whose presence is precisely what we want to avoid with our construction. A careful assessment of EWPT would be needed to establish if $\xi\simeq0.2$ is still viable in the Twin case or if instead a stronger limit applies.\label{foot}} The second set of particles are the Twin $W$'s, which are $3$ EW--neutral particles with a common mass
\begin{equation}
M_{\widetilde{W}}^2 \simeq \frac{1}{4}  \widetilde{g}_{2}^2 f^2 \cos^2\frac{V}{f}=\frac{1}{4}  \widetilde{g}_{2}^2 f^2  (1-\xi)\,.
\end{equation}
For $\widetilde{g}_{2}\sim g_{2}$ the Twin $W$'s are light, only a factor of $1\over \sqrt{\xi}$ heavier than the $W$. Finally, we do have the $21$ strong sector resonances which are all degenerate at the leading order in the $g_{1(2)}/g_\rho$ expansion because of the unbroken $SO(7)$, with a common mass $g_\rho f/2\sim m_*$. The ES couplings break the degeneracy and the $21$ resonances organize themselves into one real ${\mathbf{3}_{\mathbf{0}}}$, one complex ${\mathbf{1_{\mathbf{1}}}}$ and three ${\mathbf{2_{\mathbf{1/2}}}}$'s of the SM group, plus four real ${\mathbf{1_{\mathbf{0}}}}$ singlets with masses 
\begin{eqnarray}
&M_{{\mathbf{3}_{\mathbf{0}}}}^2\simeq \frac{1}{4} f^2 (g_\rho^2+g_{2}^2) \,,\;\;M_{{\mathbf{1}_{\mathbf{1}}}}^2\simeq \frac{1}{4} f^2 g_\rho^2 \,,\;\; M_{{\mathbf{2}_{\mathbf{1/2}}}}^2\simeq \frac{1}{4} f^2 g_\rho^2 \,,
&\nonumber\\
&
M_{{\mathbf{1}_{\mathbf{0}}},1}^2\simeq \frac{1}{4} f^2 (g_\rho^2+g_{1}^2) \,,\;\;
M_{{\mathbf{1}_{\mathbf{0}}},2}^2\simeq \frac{1}{4} f^2 (g_\rho^2+\widetilde{g}_{2}^2) \,.&
\end{eqnarray}
Notice that many of the composite resonances are charged under the EW group, unlike the elementary Twin $W$'s which are EW--singlets, and thus they could be directly produced at the LHC at a significant rate. However their coupling to SM fermions rapidly decrease for increasing $g_\rho$ making current limits on their mass safely below $2$~TeV already for $g_\rho\gtrsim2$ \cite{Pappadopulo:2014qza}. The leading constraint comes from their contribution to the $\hat{S}$ parameter of EWPT, which places them above $2$ or $3$~TeV \cite{Barbieri:2004qk}. This threshold corresponds, for $f=500$~GeV, to a large but still reasonable coupling $g_\rho\simeq g_*\sim 6$. The spin--one particle spectrum of our construction, summarized in the left panel of Figure~ \ref{spec}, displays the typical pattern of Twin Higgs models.

The advantage of a $2$--site model is that it makes the Composite Higgs potential calculable at one loop up to logarithmic divergences. The potential arises from loops of the  ES, which, as explained above,  breaks of the Goldstone symmetry. Focusing momentarily on the loops of the SM $W$'s and of their Twin partners, and working at the leading order in the $g_{2}/g_\rho$ expansion we obtain 
\begin{equation}
\label{pg22}
V_{g_2^2}[H] = \frac{9 g_\rho^2 f^4}{512\pi^2}  \left( g_{2}^2\sin^2\frac{H}{f}+  \widetilde{g}_{2}^2\cos^2\frac{H}{f}\right) \left(1+\log \frac{4 \mu ^2}{g_\rho^2f^2 }\right)\,.
\end{equation}
The logarithmic term in the equation stems for the previously--mentioned divergent contribution to the potential, which will be cut--off at the scale $\mu$ where other CS resonances, not included in our description, appear. Given that we expect those not to be far, we will not take this logarithm seriously and treat it as order one in our estimates.\footnote{The potential could be made fully calculable with a $3$--site model \cite{Panico:2011pw} and no large logarithm would appear in this case barring an unnatural separation among the two layers of resonances.}

What is remarkable and non--generic in Eq.~(\ref{pg22}) is that for $g_{2}$ exactly equal to $\widetilde{g}_{2}$ the $\sin^2$ and $\cos^2$ terms sum up to $1$ and the potential becomes an irrelevant shift of the vacuum energy. This result is compatible with the original Twin Higgs argument \cite{Chacko:2005pe}, according to which the quadratically divergent contributions to the Higgs potential, of order $g^2f^2\Lambda^2/16\pi^2$, cancel in the Twin--symmetric limit $g=\widetilde{g}$. Given that from the low--energy perspective of Ref.~\cite{Chacko:2005pe} the cutoff $\Lambda$ is the resonance scale $m_*\simeq g_\rho f$, this is precisely what we are finding here. However the true reason that  underlies the cancellation is slightly different and we believe it is  important to clarify this conceptual point. This also has a practical implication we will describe below. 

The functional form of the potential in Eq.~(\ref{pg22}) can be obtained by spurion analysis, with the method developed in \cite{Mrazek:2011iu}, by assigning $G$ quantum numbers to the Elementary/Composite couplings which break the Goldstone symmetry. The ES couples via gauging to the Composite one, {\it{i.e.}} by mixing with the corresponding global current operators. By focusing on the $W$ and $\widetilde{W}$ interactions, which are the ones responsible for the potential (\ref{pg22}), these can be written as
\beq
{\mathcal{L}}_{\textrm{INT}}=g_{2} W_\mu^\alpha ({J_{L}})^\mu_\alpha + \widetilde{g}_{2} \widetilde{W}_\mu^\alpha (\widetilde{J}_{L})^\mu_\alpha \,,
\eeq
where $J_{L}$ and $\widetilde{J}_{L}$ are the currents associated with the generators $T_L$ and $\widetilde{T}_L$. With respect to our previous notation here we performed a field redefinition $W\rightarrow g_{2} W$ and $\widetilde{W}\rightarrow \widetilde{g}_{2} \widetilde{W}$ to move  the couplings from the kinetic term to  the interaction terms. We can then uplift the  couplings to two spurions ${G}^A_\alpha$ and $\widetilde{{G}}^A_\alpha$ with an index $A$ in the ${\mathbf{28}}$ of $SO(8)$ and an index $\alpha=1,2,3$, so as to rewrite ${\mathcal{L}}_{\textrm{INT}}$ in a formally invariant fashion 
\beq
{\mathcal{L}}_{\textrm{INT}}= W_\mu^\alpha {G}^A_\alpha ({J_L})^\mu_A + \widetilde{W}_\mu^\alpha  {\widetilde{G}}^A_\alpha ({\widetilde{J}_{L}})^\mu_A \,.
\eeq
The two spurions are identical from the viewpoint of the CS and thus they enter  the potential in  exactly the  same way. What makes them different is the physical values to which we will eventually set them. By switching to a matrix notation we have 
\begin{equation}
G_\alpha\equiv {G}^A_\alpha T^a = g_{2} T_{\bf L}^\alpha\,, \qquad {\widetilde{G}}_\alpha \equiv {\widetilde{G}}^A_\alpha T^A= \widetilde{g}_{2} \widetilde{T}_{\bf L}^\alpha\,.
\end{equation}

Finding the structures that can appear in the potential at order $g_{2}^2$ and $\widetilde{g}_2^2$ amounts to classifying the $G$-invariants that can be constructed with two of those spurions and the Goldstone Matrix in Eq.~(\ref{golmat}). It has been shown in Ref.~ \cite{Mrazek:2011iu} that the number of independent invariants is equal to the number of singlets of the unbroken group $H$ that can be obtained out of the various spurion components, minus the number of singlets of the full group $G$. In the present case the spurions are in the Adjoint of $G=SO(8)$, which decomposes as ${\mathbf{28}}={\mathbf{21}}\oplus{\mathbf{7}}$ under $H=SO(7)$. Since one $SO(7)$ singlet is present in the product of two ${\mathbf{21}}$'s and one in the product of two ${\mathbf{7}}$'s, but one full $SO(8)$ singlet arises from two ${\mathbf{28}}$'s, only one invariant exists, given by
\beq
I=\sum_{\alpha,\hat{a}}\left\{{\textrm{Tr}}[T_{\mathbf{7}}^{\hat{a}} U^t {G}_\alpha U]\right\}^2\,.
\eeq
Depending on which of the physical spurions is inserted, we obtain a different dependence on the Higgs field
\begin{equation}
I =  \frac{3}{4} g_{2}^2\sin^2\frac{H}{f}\,,\;\;\;\;\; \widetilde{I} =  \frac{3}{4} \widetilde{g}_{2}^2\cos^2\frac{H}{f}\,.
\end{equation}
The two spurions are treated by the CS in exactly the same way, therefore the two terms above must appear in the potential with the same coefficient. That explains the form of  Eq.~(\ref{pg22}) and originates  the cancellation at $g_2=\widetilde{g}_2$.

The above argument  is based on the symmetries and the selection rules of the underlying UV theory and is thus  completely conclusive. That is instead not the case of the original Twin Higgs reasoning, which only establishes the cancellation of quadratic divergences. The reason why this could not be enough is that the quadratic divergence corresponds, from the UV viewpoint, only to some of the contributions to the potential, namely the ones coming from the high--scale propagation of the light degrees of freedom. The effects of heavy resonances are equally sizable and they cannot be controlled by a purely low--energy ``calculation'' of the quadratic divergence. One might thus expect that in some situations the quadratic divergence might cancel in the low--energy theory, but still equally large finite contributions arise in the complete models making the Twin Higgs cancellation ineffective. One example of that  is provided by the non--custodial Twin Higgs model, based on the $SU(4)/SU(3)$ coset where the $W$ and their Twins gauge the $SU(2)\times \widetilde{SU}(2)$ subgroup. As we  explicitly verified  the cancellation does not occur in a $2$--site implementation of this scenario, meaning that order $g^2f^2m_\rho^2/16\pi^2$ term are present also in the Twin--symmetric limit and should be taken into account in the study of the potential. A straightforward spurion analysis offers a simple criterion to understand under what condition the quadratic divergence argument will either fail, as in the $SU(4)/SU(3)$ case, or be   uplifted to a proper selection rule, as in the case of  $SO(8)/SO(7)$. The point is that the quadratic divergence contribution to the potential itself does respect the symmetries and the selection rules of the theory, and therefore it must have a functional form which is allowed by the spurion analysis. In $SO(8)/SO(7)$ there is only one invariant, and thus the $g^2$ and $\widetilde{g}^2$ terms in the quadratic divergence must have the same functional dependence on the Higgs VEV as the corresponding terms in the full potential. If from the low--energy calculation we find that they have the appropriate form to cancel, for instance a $\sin^2$ plus $\cos^2$ structure, the same must occur for the complete potential. The $SU(4)/SU(3)$ Twin Higgs fails because two independent invariants exist. The naive quadratic divergence is proportional to one invariant, for which the cancellation occurs, but also the other invariant arises in general in the complete potential.\footnote{An argument showing  that  $SO(8)$ is sufficient  in order to fully protect the Higgs mass at $O(g^2)$ can also  be found in Appendix~B of Ref.~\cite{Chacko:2005un}. Freed of inessential details, the argument can be synthesized as follows. Under the $SU(4)\times U(1)$ subgroup of $SO(8)$, the adjoint and fundamental irreps of $SO(8)$ decompose respectively as ${\bf 28}={\bf 1_0}+{\bf 6_2}+{\bf 6_{-2}}+{\bf 15_0}$ and ${\bf 8}={\bf 4_1}+{\bf \bar 4_{-1}}$. Each different generator of $SU(2)_L\times \widetilde{SU}(2)_L$  with definite twin parity ($T_L^a\pm\widetilde T_L^a$) transforms as the singlet ${\bf 1_0}$ of a different $SU(4)\times U(1)$ subgroup. In the twin symmetric limit, $g=\widetilde g$, the vector bosons associated with the above twin parity eigenstates are also propagation eigenstates and  the $O(g^2)$ correction to the effective action can  be written as the sum over single exchanges of such eigenstates. Therefore each such contribution  respects a different $SU(4)\times U(1)$. Now,  $SU(4)\times U(1)$ invariants built from the submultiplets of the  ${\bf 8}$ of $SO(8)$ accidentally respect the full $SO(8)$. As the Goldstone bosons of $SO(8)\to SO(7)$ can be made to live inside the $\bf 8$ of $SO(8)$, we conclude that at  $O(g^2)$ the potential respects $SO(8)$ and thus the Goldstone bosons remain massless. While  the above argument is not unrelated to our derivation, we find it specific to that particular case. We think our methodology, based on the analysis of the invariants constructed with ``Goldstone-dressed" external couplings, is both  more systematic, encompassing in particular fermionic couplings, and more direct. For instance, it immediately   outlines
 the structural difference between $SU(4)$ and $SO(8)$, which was in fact not appreciated in Ref.~\cite{Chacko:2005un}.}  

The reader might wonder at this point what is the role of the Twin Parity symmetry in our discussion. It actually played no role up to now, but it becomes essential when trying to really realize the cancellation via the condition $g_{2}=\widetilde{g}_{2}$. This can be enforced by Twin Parity, which is defined as the operation
\beq
W_\mu\;\leftrightarrow\;\widetilde{W}_\mu\,,
\eeq
which flips the $W$'s with their Twin partners, supplemented by a transformation on the CS which interchanges the $SO(4)_L$ and $\widetilde{SO}(4)_L$. The latter is an element of $SO(8)$, 
\beq
\label{partwin}
{\mathcal{P}}_{\textrm{Twin}}=\left[\begin{array}{cc}
0 & \Id_4\\
\Id_4 & 0
\end{array}\right]\,,
\eeq
and thus it is automatically a symmetry of our construction.

An exact Twin symmetry requires $g_{2}=\widetilde{g}_{2}$, but it would also require the existence of a Twin partner of the Hypercharge gauge boson, which however we have not introduced. Twin Parity is thus broken by the Hypercharge and thus in the Higgs potential we find an unsuppressed $g_{1}^2$ contribution of the form
\begin{equation}
\label{pg2}
V_{g_1^2} = \frac{3 g_\rho^2 f^4}{512\pi^2}  g_{1}^2\sin^2\frac{H}{f} \left(1+\log \frac{4 \mu ^2}{g_\rho^2f^2 }\right)\,.
\end{equation}

\subsection{The fermionic sector}

To understand the symmetry breaking potential it is crucial to describe properly the source of the top mass. It originates, as in the canonical Composite Higgs, from a linear interaction among the elementary top fields and some Composite Sector fermionic operators. This realizes the so--called ``Partial Compositeness'' paradigm  \cite{Kaplan:1991dc}. The low--energy description of the setup depends on the choice of the quantum numbers of the latter fermionic operators under the CS global group. Here we take the elementary $q_L$ doublet to interact with an $\bold{8}$ of $SO(8)$ and the elementary $t_R$ to interact with a singlet operator. This choice is not only simple and minimal, it is also suited to discuss the case of a composite $t_R$ field, as we will see below.

Adding fermions requires, again as in the ordinary Composite Higgs, the presence of additional unbroken global symmetries of the CS. In the first place, a $q_L$ doublet with $1/6$ Hypercharge does not fit in an $\bold{8}$ if the Hypercharge is completely internal to the $SO(8)$ group. We will thus consider a global $U(1)_X$, define  Hypercharge as  $Y=T_{{\bf R}}^3+ X$ and assign appropriate $U(1)_X$ quantum numbers to our fields. Second, and more importantly, the  $SU(3)_c$ color group of QCD must be assumed to be an unbroken symmetry of the CS. This is because the quarks are color triplets and thus the CS must carry QCD color to interact linearly with them. Clearly there is additional structure in the Twin Composite Higgs. First of all, a second set of ES doublet and singlet fields $\widetilde{q}_L$ and $\widetilde{t}_R$ are introduced and coupled to an  $\bold{8}$  and to a singlet of $SO(8)$, respectively. We call these particles the ``Twin Partners'' of the Top (and $b_L$) quarks. Second, since we do not want them to be colored or charged under any of the SM groups but still we want them to be related by a symmetry to $q_L$ and $t_R$, also Twin $\widetilde{U}(1)_X$ and Twin $\widetilde{SU}(3)_c$ color global groups have to be introduced.

Let us now turn to our model, which incorporates fermions by a standard $2$--site construction \cite{Panico:2011pw}. The spirit is again to describe a minimal set of CS resonances, compatible with the structure of the underlying CS theory. Given that we assumed the elementary $q_L$ to be coupled to one fermionic operator in the $\bold{8}$ of $SO(8)$, which decomposes under the unbroken $SO(7)$ as  $\bf 8 = \bf 7 \oplus 1$, it is reasonable to expect a $\bf 7$ and a singlet of fermionic resonances in the spectrum, namely
\begin{equation}
\Psi = \left( \begin{array}{ll}
\Psi_{\bf 7}\\
\Psi_{\bf 1}
\end{array}\right),
\end{equation}
The operators, and consequently the associated resonances, must be in a color triplet and  must carry  $U(1)_X$ charge $2/3$ in order to couple to $q_L$. Similar considerations hold for the $t_R$, which mixes with a singlet operator with $X=2/3$. This suggests the existence of a singlet, which however we have already incorporated by the field $\Psi_{\bf 1}$. The $\Psi$ resonances are the so--called ``Top Partners'', they carry QCD color as in the ordinary Composite Higgs scenario. However in the Twin Higgs case naturalness will place a weaker bound on their mass. Identical considerations hold for the  Twin Tops and their couplings to the CS, which suggest the existence of a second set of fermionic resonances
\begin{equation}
\widetilde{\Psi} = \left(
\begin{array}{ll}
\widetilde{\Psi}_{\bf 7} \\
\widetilde{\Psi}_{\bf 1}
\end{array}\right)\,.
\end{equation}
Those are once again a $\bf 7$ and a singlet of $SO(7)$, but they are not identical to the untilded $\Psi$ because they are neutral under the ordinary color and $U(1)_X$ while they are charged under the Twin $\widetilde{SU}(3)$ and $\widetilde{U}(1)_X$.

The decomposition of the Top Partners $\Psi$ and their Twins $\widetilde{\Psi}$ into SM representations is described in Appendix~\ref{CompMult}. As far as $\Psi$ is concerned, its $8$ components decompose under the standard electroweak gauge group into one ${\mathbf{2}}_{\mathbf{1/6}}$ and one ${\mathbf{2}}_{\mathbf{7/6}}$ plus four states in the ${\mathbf{1}}_{\mathbf{2/3}}$. The phenomenology of these particles is expected to be similar to that of the Top Partners in the ordinary Composite Higgs model\cite{DeSimone:2012fs}. The eight components of the Twin $\widetilde{\Psi}$'s decompose into a ${\mathbf{2}}_{\mathbf{1/2}}$, a ${\mathbf{2}}_{\mathbf{-1/2}}$  and four neutral singlets ${\mathbf{1}}_{\mathbf{0}}$. Unlike the $\Psi$'s, they carry no QCD color but some of them still  communicate directly with the SM by EW interactions.

Now that the field content has been specified, we can write down our Lagrangian. Leaving aside the kinetic terms, the gauge interactions and the couplings of the fermions with the vector resonances which will not play any role in what follows, we have 
\begin{equation}
\begin{array}{ll} 
\label{lagtop}
\mathcal{L}_{top} = & \displaystyle
\left[ y_L f (\bar{Q}_L)^I\Sigma_{I i}(\Psi_R)^{i} + \widetilde{y}_L f (\bar{\widetilde{Q}}_L)^I\Sigma_{I i}(\widetilde{\Psi}_R)^{i} + \right.\\
& \displaystyle \left.  {y_R } f \bar{t}_R \Psi_{L{\bf 1}}  + {\widetilde{y}_R } f \bar{\widetilde{t}}_R \widetilde{\Psi}_{L{\bf 1}} + \text{ h.c. }\right] \\
& \displaystyle  - M_\Psi \bar{\Psi}_{\bf 7}\Psi_{\bf 7} - \widetilde{M}_\Psi \bar{\widetilde{\Psi}}_{\bf 7}\widetilde{\Psi}_{\bf 7}  - M_{S} \bar{\Psi}_{\bf 1} \Psi_{\bf 1} - \widetilde{M}_{S} \bar{\widetilde{\Psi}}_{\bf 1} \widetilde{\Psi}_{\bf 1}\,,
\end{array}
\end{equation}
where the Elementary $q_L$ and its mirror are embedded into incomplete octets 
\begin{equation}
Q_L= \frac{1}{\sqrt{2}}\left( \begin{array}{c}
i b_L \\
b_L \\
i t_L \\
-t_L \\
0 \\
0 \\ 
0 \\
0 \\
\end{array}
\right)\,,\;\;\;\;\;
(\widetilde{Q}_L)^I = \frac{1}{\sqrt{2}}\left( \begin{array}{c}
0 \\
0 \\ 
0 \\
0 \\
i \widetilde{b}_L \\
\widetilde{b}_L \\
i \widetilde{t}_L \\
-\widetilde{t}_L \\
\end{array}
\right)\,.
\end{equation}
Notice that the ES fields, compatibly with the Partial Compositeness hypothesis, are taken to interact linearly with the CS through mass--mixings with the resonance fields. The non--vanishing entries of the embeddings $Q_L$ and $\widetilde{Q}_L$ are of course precisely designed to make $q_L$ and $\widetilde{q}_L$ couple to components of $\Psi$ and $\widetilde{\Psi}$ with the appropriate gauge quantum numbers. The couplings  $y_L$ and $\widetilde{y}_L$ control the strength of the interaction between Elementary and Composite fermions and are assumed to be weak,
 namely $y_L,\widetilde{y}_L\ll g_*$. The mass parameters $M_\Psi (\widetilde{M}_\Psi)$ and $M_S(\widetilde{M}_S)$ come instead purely from the CS. We thus expect them to be of order $m_*$, around  the scale of the vector resonances described in the previous section. As far as the $t_R$ and $\widetilde{t}_R$ mixing are concerned, two interpretations are possible which lead to different estimates for the size of the associated couplings $y_R$ and $\widetilde{y}_R$. If we regard $t_R$ and $\widetilde{t}_R$ as ES fields, the couplings have to be weak, much below $g_*$ and possibly close to their left--handed counterparts. However we can also interpret  $t_R$ and $\widetilde{t}_R$ as completely composite chiral bound states originating from the CS, perhaps kept exactly massless by some anomaly matching condition. If it is so, their mixing is a purely CS effect and thus $y_R,\widetilde{y}_R\sim g_*$. We will consider both options in what follows taking also into account the possibility of smoothly interpolating between the two.

As a part of the Composite Twin Higgs construction we do have to impose Twin Parity, at least to some extent as described in the previous section. Twin Parity acts as 
\beq
Q_L\leftrightarrow\widetilde{Q}_L\,,\;\;t_R\leftrightarrow\widetilde{t}_R\,,\;\;\Psi\leftrightarrow\widetilde{\Psi}\,,
\eeq
times the $SO(8)$ transformation in Eq.~(\ref{partwin}) acting on the resonance fields $\Psi$ and ${\widetilde{\Psi}}$.\footnote{We have not mentioned the mirror gluons which gauge $\widetilde{SU}(3)_c$, needless to say they also get exchanged with the SM gluons.} If it were an exact symmetry it would imply all masses and couplings in the Lagrangian (\ref{lagtop}) to be equal to their Twin, un--tilded,  counterparts. We notice that the implementation of Twin Parity is slightly different in the fermionic and  gauge sectors. In the gauge sector of the CS, Twin Parity was acting just like an $SO(8)$ transformation and thus it was automatically a symmetry. Now instead Twin Parity entails the exchange of different fermionic CS resonances, charged under different global groups. Imposing Twin Parity thus becomes a non--trivial constraint on the CS.

We can now turn to the determination of the mass spectrum.  By working in the limit $y_L,{\widetilde{y}}_L\ll g_*$, we will focus  on the leading relavant order in an expansion in powers of $y_L$ and ${\widetilde{y}}_L$.  We will instead not treat $y_R$ and ${\widetilde{y}}_R$ as small parameters, so that our formulae will hold for both completely composite and partially elementary right--handed fields. Aside from the exactly massless $b_L$ and $\widetilde{b}_L$ --which will get a mass by mixing with other resonances or by some other unspecified mechanism--, the lightest particles are the Top quark and its Twin partner, with masses
\begin{equation}
\label{ttpmass}
\begin{array}{ll}
\displaystyle m_{t}^2 \simeq {f^4 \over 2} \frac{y_L^2 y_R^2}{ M_S^2+  y_R^2 f^2} \xi\,, \qquad \displaystyle m_{\widetilde{t}}^2 \simeq {f^4 \over 2} \frac{\widetilde{y}_L^2 \widetilde{y}_R^2}{\widetilde{M}_S^2+ \widetilde{y}_R^2 f^2} (1-\xi)\,.
\end{array}
\end{equation}
If we remember that $M_S\sim m_*=g_*f$ and $y_Rf$ is either $\sim m_*$ or smaller for a partially elementary $t_R$, we see that the Top mass respects the usual Partial compositeness estimate
\beq
\label{partcomest}
m_t=\frac{y_t}{\sqrt{2}} \cdot v\sim \frac{y_L y_R}{g_*}\cdot v\,,
\eeq
out of which we can determine the size of  $y_L$ in terms of the other parameters. If $t_R$ is completely Composite, we expect $y_R\sim g_*$ and thus $y_L$ must be around the physical Top Yukawa coupling $y_t\sim1$. Larger values are obtained in the case of a partially Elementary $t_R$. The same parametric estimate can be performed for the Twin Top, whose mass scales like
\beq
m_{\widetilde{t}}\sim \frac{\widetilde{y}_L \widetilde{y}_R}{g_*}\cdot f\,.
\eeq
Differently from the Top one, the Twin Top mass is not proportional to $v$ but to $f$ because the Twin $\widetilde{SU}(2)_L$ is broken by the CS directly at the scale $f$. 

The rest of the spectrum comprises the $16$   components of $\Psi$ and $\widetilde{\Psi}$. They all have masses of order $m_*$, though not degenerate because of the freedom to choose the CS mass parameters $M_\Psi\neq M_S$, ${\widetilde{M}}_\Psi\neq {\widetilde{M}}_S$. We expect two almost degenerate $7$--plets, with mass $M_\Psi$ and $\widetilde{M}_\Psi$ respectively, plus $2$ singlets whose masses are controlled by $M_S$ and $\widetilde{M}_S$ and by the $y_R f$ and $\widetilde{y}_R f$ mixings. The interaction with $q_L$ and $\widetilde{q}_L$ remove part of the degeneracy and the spectrum organizes in degenerate SM multiplets as described above, with splitting of order $y_L^2f^2$ and $\widetilde{y}_L^2f^2$ in the mass squared. Further tiny splitting emerge after EWSB. The qualitative structure of the spectrum respects the Twin Higgs expectation depicted in the right panel of Figure~\ref{spec}.

Let us finally turn to the calculation of the Higgs potential, working once again in the weak coupling expansion $y_L,\widetilde{y}_L\ll g_*$. Notice that $y_L$ and $\widetilde{y}_L$ are the only sources of $SO(8)$ breaking in our fermionic Lagrangian, therefore the Higgs potential must be proportional to powers of those couplings. It receives its formally leading contribution at  second order in the coupling expansion, through a term
\begin{eqnarray}
&\displaystyle V_{y^2}(H) = \frac{N_c f^2}{32 \pi ^2} & \left\{y_L^2 \left[M_{\Psi }^2 \log\frac{\mu^2}{M_\Psi^2} - M_S^2 \log\frac{\mu^2}{M_S^2+{f^2 y_R^2}} \right]\cdot\sin^2 \frac{ h}{f}\right.\nonumber\\
&\ &\left.+\widetilde{y}_L^2 \left[\widetilde{M}_{\Psi }^2 \log\frac{\mu^2}{\widetilde{M}_\Psi^2}-\widetilde{M}_S^2 \log\frac{\mu^2}{\widetilde{M}_S^2 + {f^2 \widetilde{y}_R^2 }} \right]\cdot\cos^2 \frac{ h}{f}\right\}\,.
\end{eqnarray}
Again, as in the order $g_2^2$ potential in the previous section, we see the Twin Higgs cancellation mechanism at work. If Twin Parity is exact so that tilded and un--tilded quantities are equal, the $\sin^2$ and $\cos^2$ sum up to one and no contribution is left to the Higgs potential. As in the gauge sector this cancellation can be explained in terms of symmetries and selection rules.  The relevant spurions in this case are the Elementary $q_L$ and $\widetilde{q}_L$ couplings, which transform in the $\mathbf{8}$ of $SO(8)$. Only one non--trivial invariant can be formed out of two $\mathbf{8}$'s,and that precisely takes the $\sin^2$ and $\cos^2$ forms of the equation above. 

The second relevant term in the potential is due to an IR effect. By looking at the spectrum of the theory in Figure~\ref{spec} we see that there is a considerable gap among the Top Partner scale $m_*=g_*f$ and the Top plus its Twin, with masses of order $y_L v$ and $\widetilde{y}_L f$. The low--energy Higgs potential thus receives a considerable log--enhanced contribution that corresponds to the  RG evolution of the Higgs quartic coupling down from the scale $m_*$.  
In our model, the well known effect of the Top is complemented by the effect of its Twin, so that
 the potential reads
\beq
\displaystyle
V_{IR}(H)=\frac{N_c}{16 \pi ^2}\left[{m_t(H)^4 } \log {m_*^2 \over m_t(H)^2}
+{m_{\widetilde{t}}(H)^4 } \log {m_*^2 \over m_{\widetilde{t}}(H)^2}
\right]\,,
\eeq
where $m_t(H)$ and $m_{\widetilde{t}}(H)$ are the Higgs--dependent Top and Twin Top masses which we can extract from Eq.~(\ref{ttpmass}). They can be expressed as 
\beq
m_t(H)^2=\frac{y_t^2}{2}f^2\sin^2\frac{H}{f}\,,\;\;\;\; m_{\widetilde{t}}(H)^2=\frac{y_{\widetilde{t}}^2}{2}f^2\cos^2\frac{H}{f}\,,
\eeq
in terms of the physical Top Yukawa and its Twin
\beq
y_t^2= \frac{\, y_L^2 y_R^2 f^2}{ M_S^2+  y_R^2 f^2}\,,\;\;\;\;\;y_{\widetilde{t}}^2= \frac{\, {\widetilde{y}}_L^2 {\widetilde{y}}_R^2 f^2}{{\widetilde{M}}_S^2+  {\widetilde{y}}_R^2 f^2}\,.
\eeq
This allows to rewrite the IR potential in an explicit form
\beq
\displaystyle
V_{IR}(H)=\frac{N_cf^4}{64 \pi ^2}\left[y_t^4 \sin^4\frac{H}{f} \log {{2m_*^2 } \over {y_t^2 f^2\sin^2\frac{H}{f} }}
+y_{\widetilde{t}}^4 \cos^4\frac{H}{f} \log {{2m_*^2 } \over {y_{\widetilde{t}}^2 f^2\cos^2\frac{H}{f} }}
\right]\,.
\eeq
Notice that an analogous IR term plays an important role in the Higgs dynamics of the MSSM with heavy stops, and so it will in our case.

The last term which we have to discuss is the contribution purely of order $y^4$, not enhanced by any IR log. The resulting expression is complicated and it will not be reported here, what matters is that it has the parametric form
\beq
\label{y4}
\displaystyle V_{y^4}(H)=\frac{N_cf^4}{128 \pi ^2}\left[(y_L^4F_1+\widetilde{y}_L^4\widetilde{F}_1)\left(\sin^4\frac{H}{f}+\cos^4\frac{H}{f}\right)+
(y_L^4F_2-\widetilde{y}_L^4\widetilde{F}_2)\left(\sin^2\frac{H}{f}-\cos^2\frac{H}{f}\right)\, .
\right]\,,
\eeq
Here $F_1$, $F_2$  are $O(1)$ functions of the mass ratios $M_S/M_\Psi$ and $y_R/M_\Psi$. The same comment applies to the corresponding tilded quantities.  The coefficient in the first parenthesis is even under the exchange of tilded with un--tilded objects, while the second one is odd and thus vanishes for exact Twin Parity. 

\section{Electroweak symmetry breaking}
 
Let us now discuss if and under what conditions we can achieve a realistic vacuum dynamics in our model. That amounts to producing electroweak symmetry breaking, the correct Higgs mass and  a sufficiently small (tunable) value of the ratio $\xi =v^2/f^2$, which controls Higgs couplings and precision electroweak observables. In the spirit of Twin Higgs, and differently from ordinary Composite Higgs models, we would like to obtain that without the need of relatively light Top Partner(s) close to the Goldstone scale $f$. Namely, we would like to keep $M_\Psi/f\equiv g_\Psi\sim g_*$ large and possibly close to the perturbativity bound $g_*\sim4\pi$. 

Let us consider first the exact Twin Parity limit, in which the untilded and tilded parameters are taken to coincide and moreover the SM hypecharge coupling $g_1$ is set to vanish. Remember indeed that in our proposal the Twin Hypercharge is not gauged and thus the SM Hypercharge gauging breaks Twin Parity. The potential, as computed in the previous section, can be written as 
  \begin{equation}
 V^{sym.}(H) = f^4 \beta\left( s^4\log\frac{a}{s^2} + c^4\log\frac{a}{c^2}\right), 
 \label{potsymm}
 \end{equation}
 where
   \begin{equation}
 s^2 \equiv \sin^2\frac{H}{f} ,~~ c^2  \equiv \cos^2\frac{H}{f}\,,
 \end{equation}
  \begin{equation}
  \label{beta}
 \beta=\frac{3 y_t^4}{64\pi^2}\,,
 \end{equation}
 and
 \begin{equation}
\log a=\log \frac{2\mu^2}{y_t^2 f^2}+\frac{y_L^4}{y_t^4}F_1\,,
 \label{quartic}
  \end{equation}
where $F_1$, which coincides with $\widetilde{F}_1$ in the Twin symmetric case, was introduced in Eq.~(\ref{y4}). This potential is not realistic. For $\log a>3/2-\log{2}$ it is minimized at the Twin symmetric point $s=c=1/\sqrt 2$, while for $\log a<1/2$ it has Twin breaking minima at respectively $s=0$, $c=1$ and $s=1$, $c=0$. In the intermediate range  $1/2< \log a<3/2-\log{2}$ it does have 
a tunable minimum with $c\not =s\not = 0$: when $\log a $ approaches $1/2$ from the above,
 $\xi$ approaches $0$. However the effective Higgs quartic in this case is purely generated by RG evolution in the SM, and it results too small unless $f\gsim 10^{10}$ GeV, which we find unacceptable from the stanpoint of fine tuning.
In conclusion none of the above cases corresponds to a realistic phenomenology.
 
A realistic potential can only be obtained by turning on the Twin Parity breaking  sources. We think a consistent picture can be obtained by treating Hypercharge as the main source of that breaking. Its effects  can be classified by the loop order at which they arise. At one loop there is the gauge contribution in Eq. (\ref{pg2}). That equation features a logarithmic divergence, but in a realistic model, that logarithm would be saturated at the scale of the strong resonances: $\mu \to m_*$.
However, known theorems fix the sign of that contribution to the potential to always be positive. That is indeed compatible with the leading log behaviour at $\mu\gg m_\rho$ in Eq. (\ref{pg2}). Another source of breaking is the Hypercharge contribution to the  RG evolution  of the top sector parameters, down to $m_*$ from the  UV scale $\Lambda_{UV}\gg m_*$, where our model is 
microscopically defined. In general this RG contribution may turn on several effects in the composite sector. In particular each and every Yukawa and mass parameter in the top sector can be affected. However under the assumption that the composite sector does not possess any twin-parity-odd relevant or marginal operator, the only couplings that will be affected are the elementary-composite mixings $y_L$ and potentially, if $t_R$ is Elementary, $y_R$. Focusing on $y_L$, which affects the potential, we expect  RG evolution to  generate
a twin breaking splitting (for the couplings renormalized at the scale $m_*$) of the form
\begin{equation}
 y_L^2 - \widetilde y_L^2 = \frac{b g_1^2}{16\pi^2}y_L^2 \log{\frac{\Lambda_{UV}}{m_*}}\equiv \Delta y_L^2
 \label{ysplit}
 \end{equation}
 where $b$ is an unpredictable numerical coefficient of order unity. In principle if the strong sector between $\Lambda_{UV}$
 and $m_*$ is approximately conformal, $b$ could be related to the OPE coefficients performing conformal perturbation theory.
 In the case of perturbative theories, where the mixing is simply provided by mass terms, we know that  $b>0$.
 That is the well known sign of the running of masses induced by gauge interactions: it makes $y_L$ grow when running towards the IR, and does not affect $\widetilde y_L$ as it involves hypercharge neutral states.
 Although we have not studied the problem, we suspect  $b>0$ is a robust feature also at strong coupling, though we shall not strongly rely on that. The insertion of Eq.~(\ref{ysplit}) in the fermion induced 1-loop potential will give rise to a two-loop contribution
 enhanced by the UV log. 
 We should also notice that analogous effects are induced on the $SU(3)$ and $SU(2)$ gauge couplings but they are numerically irrelevant.

The net effect of all the above considerations is the addition to the potential in Eq.~(\ref{potsymm}) of a Twin breaking term
\begin{equation}
\Delta V(H) = \alpha f^4s^2
\end{equation}
\begin{equation}
\alpha = \frac{3 g_1^2 g_\rho^2 }{512\pi^2} A
+\frac{3 \Delta y^2 g_\Psi^2}{32\pi^2} B,
\end{equation}
where $g_\Psi=M_\Psi/f$ is the effective coupling associated with the overall size of the fermion masses introduced above --which we expect to be of order $g_*$-- and $g_\rho$ is the vector coupling, which is also expected to be around $g_*$. Finally $A$ and $B$ are numerical coefficients that depend on the details of the model. $A$, as we mentioned, is robustly predicted to be positive, while $B$ can take either sign. 

 The overall potential 
 \begin{equation}
{ V(H) \over f^4} = \alpha  s^2 + \beta\left( s^4\log\frac{a}{s^2} + c^4\log\frac{a}{c^2}\right)
 \label{simplepot}
 \end{equation}
  is now capable to give rise to the desirable  pattern of electroweak symmetry breaking. In order to achieve that, $\alpha$ must be positive. One is immediately convinced of that, by working with the non canonical field $\phi = f\sin h/f$. In this parametrization $\alpha$ only affects the quadratic part of the potential, and the quartic term $\phi^4$ purely comes from the twin symmetric contribution: a positive effective quartic of the right size can only be achieved for  $a\gg 1$. But for $a\gg1$ the twin symmetric potential contributes
 a negative $\phi^2$ term when expanded around $H=0$ and this must be compensated by  tuning against a positive $\alpha$, thus obtaining a vacuum expectation value $\langle\sin^2 H/f\rangle  =\xi\ll 1$. A value $\xi \sim 0.2$ could be sufficient to account for present bounds on the Higgs couplings (see however Footnote~\ref{foot}). 
 
 From equation (\ref{simplepot}) we can readily study the condition for having a tunable minimum with $\xi\ll 1$. The minimization of Eq.~(\ref{simplepot}) yields
 \beq
\frac{\alpha}{\beta}=-1+2\log\frac{a}{1-\xi}+2\,\xi\left[1-2 \log\frac{a}{\sqrt{\xi(1-\xi)}}\right]\,.
\label{tuning}
\eeq
On the extremum defined by the above equation the Higgs mass is
\beq
\frac{m_H^2}{v^2}=8\beta(1-\xi)\left[\log\frac{a^2}{\xi(1-\xi)}-3\right]\,.
\eeq
For a given $\xi$, the observed masses of the Higgs and of the Top, which controls $\beta$ through Eq.~(\ref{beta}),  fix then the value of $a$. Using the $\overline{\text{MS}}$ Top Yukawa coupling at the scale $v$, we have $y_t^4\sim 0.8$  in $\beta$, so that we find 
\beq
\log a\simeq 6 +\log\sqrt \xi
 \label{estimate1}
\eeq
which for a realistic $\xi \sim 0.1$ corresponds to $\log a\sim 5$.
%
%
%
Now notice that the definition of $a$ in Eq.~(\ref{quartic}) depends on $\mu$. In a reasonable model we expect this contribution to be saturated at the mass $m_*\sim g_* f$ of the composite sector. With this interpretation, the first term in Eq.~(\ref{quartic}) is 
$\sim\log (g_*/y_t)^2$. For a maximally strongly coupled theory $g_*\sim 4\pi$, this is in the right ballpark to match Eq.~(\ref{estimate1}). For smaller $g_*$, that is for lighter resonances, the remaining term in Eq.~(\ref{quartic}) can bridge the gap and produce the needed value of $\log a$. The situation in our model is reminiscent of the MSSM with  moderately large $\tan \beta$ and heavy stops. In that case the correct quartic
is produced in equal measure  by the tree level electroweak D-terms and by the top/stop renormalization of the quartic.  
In our case the electroweak D-term is basically replaced by the Twin Top contribution. One also has to pay attention not to make $\log{a}$ too large, producing a too heavy Higgs. This would tend to be the case for a considerably Elementary $t_R$. Indeed if for instance left-- and right--handed couplings were comparable, {\it{i.e.}} $y_L\sim y_R$, from Eq.~(\ref{partcomest}) we would obtain $y_L^2\sim g_\Psi y_t$ and thus too a large contribution to $\log{a}$ from the second term in Eq.~(\ref{quartic}) unless $g_\Psi<\sqrt{6}\simeq2.4$, which means relatively Light Top Partners as in the ordinary Composite Higgs scenario. Total $t_R$ compositeness, or at least a larger compositeness for the $t_R$ than for the $q_L$, is thus preferred in our scenario.

Consider now the value of $\alpha/\beta$ needed to be able to tune $\xi \ll 1$. Eq.~(\ref{tuning}) requires a sizeable value $\alpha/\beta \sim 9$. 
One can check what that relation requires given our estimate of $\alpha$. Assuming $\alpha$ is dominated by the 1-loop IR dominated effect
implies
\begin{equation}
A \,\frac{g_\rho^2 g_1^2}{80 y_t^4}\sim 1 
\end{equation}
which seems to require even for $g_\rho \sim 4\pi$ a sizeable $A\sim 4$, borderline but perhaps acceptable.
On the other hand assuming $\alpha$ is dominated by the RG contribution we find
\begin{equation}
\frac{Bb}{80\pi^2}\frac{y_L^2}{y_t^2}\frac{g_\Psi^2}{y_t^2} g_1^2 \log \frac{\Lambda_{UV}}{m_*}\sim 1\,.
\end{equation}
This is satisfied for completely composite $t_R$, $y_L=y_t$, when
\beq
\log \frac{\Lambda_{UV}}{m_*}\sim\frac{80\,\pi^2}{b\,B\,g_1^2}\frac{y_t^2}{g_\Psi^2}\gtrsim \frac{50}{b\,B}\,,
\eeq
{\it{i.e.}} for a large separation among the IR CS confinement scale and the UV one where it originates. Overall this seems like a plausible picture.

\section{Conclusions}

A (partial) mirroring of the particles and interactions of the SM and of the new CS may give rise to non-minimal Composite Higgs models where a minimal amount of fine tuning is needed to be consistent with current bounds and, most importantly, where there is  no new particle carrying SM charge below a few TeV. This eliminates one possible signature of Composite Higgs models, namely the production of colored Partners of the Top quark \cite{Contino:2006qr}, which need to be light in the ordinary constructions \cite{Matsedonskyi:2012ym,Panico:2012uw,DeSimone:2012fs} . The limits from the non--observation of the latter particles are currently comparable with other constraints. However they could become the strongest limit after the second run of the LHC. In that case the Composite Twin Higgs scenario might come to rescue.

A consistent picture emerges with the following salient features. First, mirroring the top Yukawa and gauge couplings is enough to render innocuous the usual quadratic divergence of the Higgs mass but does not guarantee, per se, the absence of finite but large corrections proportional to the squared mass  of the resonances carrying SM charges. Extra hypotheses, which hold automatically in our construction, are needed to uplift the divergence cancellation to a structural protection of the potential. Second, the breaking of the mirror symmetry needed to get a realistic minimum of the Higgs potential may be realized by not mirroring the weak hypercharge. This is how the potential acquires a positive squared mass term, necessary to counteract the negative term from the mirror symmetric term, quartic in the top Yukawa coupling. The cancellation between these two terms is the unavoidable tuning needed to explain the smallness of the ratio $(v/f)^2$, currently below about $0.2$, as in any Twin model. On the other hand the size of the individual terms, both quadratic and quartic, is right, without any further tuning, provided the RG evolution of the top sector parameters due to hypercharge is active already at a high UV  scale which might not be far from the GUT scale \footnote{Needless to say, without a mirror hypercharge no extra massless vector occurs in the spectrum, thus avoiding possible unpleasant cosmological consequences.}. 

We think that the phenomenology of composite twin Higgs models deserves attention. The infrared effects on the EWPT is well known since long time \cite{Barbieri:2005ri}. The search for relatively light mirror states, without SM charges, may also be possible in the next LHC run. Needless to say, to see the entire spectrum of these models in its full glory requires a Future Circular Collider in the hadronic mode.

\subsubsection*{Acknowledgments}
It is a pleasure to thank R.~Torre and J.~Serra for the useful discussion. We also thank A.~Tesi for discussions related to his forthcoming paper \cite{Tesi}. D.G. would also like to thank F.~Riva for his insightful comments. The work of R.B. is supported in part by the European Programme ``Unification in the LHC Era",  contract PITN-GA-2009-237920 (UNI\-LHC) and by MIUR under the contract 2010YJ2NYW-010. D.G. and R.R. are supported by the Swiss National Science Foundation under contract 200020-150060. A.W. acknowledges the ERC Advanced Grant
no. 267985 DaMeSyFla, the MIUR-FIRB Grant RBFR12H1MW and, together with R.R., the SNF Sinergia no. CRSII2-
141847 for support. 

\subsubsection*{Note added}
While this work was under completion, a related paper appeared that contains some level of overlap with our results, \cite{Craig:2015pha}. 

\appendix 
\section{SO(8) Generators}\label{SO(8)Gen}

In this appendix, we list the twenty-eight $SO(8)$ generators, decomposing them into irreducible representations of the $SO(7)$ subgroup, $\bf 28 = \bf 7 \oplus \bf 21$. We can compactly write the generators as:
\begin{equation}
(T_{ij})_{kl} = {i \over 2} (\delta_{ik}\delta_{jl}-\delta_{il}\delta_{jk}),
\end{equation}
with $i,j,k,l = 1, \cdots , 8$. Seven of these generators are the broken ones and they transform in the $\bf 7$ of $SO(7)$; since the global symmetry of the composite sector is broken by the vev
\begin{equation}
\text{VEV} =f (0,0,0,0,0,0,0,1)^t,
\end{equation}
we can indicate the broken generators as:
\begin{equation}
(T^{\bf {7}}_\alpha)_{\beta \gamma} =  {i \over 2} (\delta_{8\beta}\delta_{\alpha \gamma}-\delta_{8\gamma}\delta_{\alpha \beta}),
\end{equation}
with $\alpha, \beta, \gamma = 1, \cdots , 7$. The remaining generators are unbroken and they transform in the adjoint of $SO(7)$; we can collectively call them:
\begin{equation}
(T^{\bf {21}}_{\alpha \beta})_{\gamma \rho} =  {i \over 2} (\delta_{\alpha \gamma}\delta_{\beta \rho}-\delta_{\alpha \rho}\delta_{\beta \gamma}),
\end{equation}
with $\alpha, \beta, \gamma, \rho = 1, \cdots, 7$ (which obviously implies $\alpha, \beta, \gamma, \delta \neq 8$ and excludes the broken case).

By taking linear combinations, we can construct the generators that contain the $SO(4) \sim SU(2)_L \times SU(2)_R$ and $\widetilde{SO}(4) \sim \widetilde{SU}(2)_L \times \widetilde{SU}(2)_R$ subgroups. These are the ones gauged in the model and can be written as:
\begin{equation}
 (T_{\bf L})^\alpha = \left(
\begin{array}{cc}
 T_{\bf L}^\alpha & 0 \\
 0 & 0 \\
\end{array}
\right), (T_{\bf R})^\alpha = \left(
\begin{array}{cc}
T_{\bf R}^\alpha & 0 \\
 0 & 0 \\
\end{array}
\right),
(\widetilde{T}_{\bf L})^\alpha = \left(
\begin{array}{cc}
 0& 0 \\
 0 & T_{\bf L}^\alpha  \\
\end{array}
\right), (\widetilde{T}_{\bf R})^\alpha = \left(
\begin{array}{cc}
 0 & 0 \\
 0 & T_{\bf R}^\alpha\\
\end{array}
\right),
\end{equation}
where $T_{\bf L}^\alpha$ and $T_{\bf R}^\alpha$ are the $4\times 4$ generators of $SO(4)$:
\begin{equation}
(T_{\bf L, R}^\alpha)_{ij} = -{i \over 2}\left[{1 \over 2} \epsilon^{\alpha \beta \gamma} \left( \delta ^\beta_i \delta ^\gamma_j - \delta ^\beta_j \delta ^\gamma_i \right) \pm \left(\delta ^\alpha_i \delta^4_j - \delta^\alpha_j \delta^4_i  \right)\right]
\end{equation}
with $\alpha = 1, \cdots 3$ and $i,j = 1, \cdots 4$.

\section{The composite multiplets}\label{CompMult}

The heavy fermion multiplets in our model form complete fundamental representations of $SO(8)$ and decompose under $SO(7)$ as described in the main text. The first multiplet, which is colored under the SM gauge group $SU(3)$ and is charged under $U(1)_X$ with $X$-charge $2/3$, contains eight heavy fermions which are organized as follows:
\begin{equation}
\Psi_{\bf 7} = {1\over \sqrt{2}} \left( 
\begin{array}{ll}
i B - i X_{5/3} \\
B + X_{5/3} \\
i T + i X_{2/3} \\
 -T +  X_{2/3} \\
\quad \sqrt{2} S_{2/3}^1\\
\quad \sqrt{2} S_{2/3}^2 \\
\quad \sqrt{2} S_{2/3}^3 \\
\end{array}
\right) , \qquad \Psi_{\bf 1} = S_{2/3}^4.
\end{equation}
The second multiplet, colored under the twin group $\widetilde{SU}(3)$ and charged under $\widetilde{U(1)}_X$ with $\widetilde{X}$-charge $2/3$, contains another set of eight heavy fermions; they are organized in a fundamental of $SO(8)$, related to the previous representation by Twin symmetry, as follows: 
\begin{equation}
\widetilde{\Psi}_{\bf 7} = {1\over \sqrt{2}} \left( 
\begin{array}{ll}
i \widetilde{D}_{-1} - i \widetilde{D}_{1} \\
\widetilde{D}_{-1} + \widetilde{D}_1 \\
i \widetilde{D}_0^1 + i \widetilde{D}^2_0 \\
-\widetilde{D}_0^1 +  \widetilde{D}^2_0\\
\quad  \sqrt{2}\widetilde{U}^1_0 \\
\quad \sqrt{2}\widetilde{U}^2_0   \\
\quad  \sqrt{2}\widetilde{U}^3_0  \\
\end{array}
\right) , \qquad  \widetilde{\Psi}_{\bf 1} = \widetilde{U}^4_0.
\end{equation}
In this notation, it is easy to decompose these heavy particles under the SM weak gauge groups. The fermions $T$, $B$, $X_{2/3}$ and $X_{5/3}$ carry all the SM quantum numbers, both in the weak and in the color sector; they decompose into two heavy doublets, $(X_{5/3}, X_{2/3})$, with electric charges $5/3$ and $2/3$ respectively, and $(T, B)$, with electric charges $2/3$ and $-1/3$ respectively. These two doublets can be therefore identified with the usual heavy fermions that we expect to exist also in conventional composite Higgs models. The remaining components of the vector $\Psi$, the $S_{2/3}^1, \cdots, S_{2/3}^4$ fields, carry mixed quantum numbers since they participate both to the SM and the Twin sector gauge interactions. In particular, they are charged under the twin weak gauge group, but they are colored under the SM $SU(3)$ and they all have electric charge equal to $2/3$. Thus they decompose as four electrically charged singlets under the SM weak gauge group. The decomposition of the Twin vector $\widetilde{\Psi}$ under the SM is quite similar. The first four components participate to the SM weak interactions, but they carry twin color quantum numbers. They decompose into two heavy doublets under the SM weak gauge group, $(\widetilde{D}_1, \widetilde{D}_0^2)$, with electric charges $1$ and $0$ respectively, and $(\widetilde{D}^1_0, \widetilde{D}_{-1})$, with electric charges $0$ and $-1$ respectively. Finally, the fields $\widetilde{U}_0^1, \cdots, \widetilde{U}_0^4$ are charged under the Twin weak and strong gauge groups and they do not carry any electric charge. They decompose therefore as four electrically neutral singlets under the SM gauge groups. 

We conclude this Appendix by commenting the action of Twin symmetry on these two vectors of heavy fermions. This symmetry can be in general decomposed as the product of two discrete symmetries. The first one can be identified as a $Z_2$ which is external to the strong sector and that rigidly interchanges $\Psi_{\bf 7}$ with $\widetilde{\Psi}_{\bf 7}$ and $\Psi_{\bf 1}$ with $\widetilde{\Psi}_{\bf 1}$. For the singlet, this is all we need to implement the Twin symmetry and we can easily identify $\widetilde{U}_0^4$ as the Twin partner of $S^4_{2/3}$. For the remaining component in the $\bf 7$, we need to make the convolution of the external discrete symmetry with an element of the unbroken symmetry group $SO(7)$, $h(\Pi)$, so that the complete Twin symmetry takes the form:
\begin{equation}
\Psi_{\bf 7} \rightarrow h(\Pi) \widetilde{\Psi}_{\bf 7}.
\end{equation}
The matrix $h(\Pi)$ is an explicit function of the Goldstone boson fields and in general it is quite complicated to work out; we expect to have a highly non-linear relation between the heavy fields in the two representations. In the limit when the Goldstone bosons are all set to zero, however, we can find a simple expression for $h$ which we can write as follows:
\begin{equation}
h = \left( \begin{array}{ccccc}
  0 & 0 & \Id_3 \\
 0 & -1 & 0  \\
 \Id_3 & 0 & 0 
\end{array}\right).
\end{equation}
By combining the action of this matrix with the external $Z_2$, we have thus an illustrative example of the action of Twin symmetry in a simple case.

%

\begin{thebibliography}{99}  
\bibitem{Giudice:2007fh}
  G.~F.~Giudice, C.~Grojean, A.~Pomarol and R.~Rattazzi,
  JHEP {\bf 0706} (2007) 045
  [hep-ph/0703164].
  
\bibitem{Agashe:2004rs}
  K.~Agashe, R.~Contino and A.~Pomarol,
  Nucl.\ Phys.\ B {\bf 719} (2005) 165
  [hep-ph/0412089].
  
\bibitem{Chacko:2005pe}
  Z.~Chacko, H.~S.~Goh and R.~Harnik,
  Phys.\ Rev.\ Lett.\  {\bf 96} (2006) 231802
  [hep-ph/0506256].
  
\bibitem{Geller:2014kta}
  M.~Geller and O.~Telem,
  arXiv:1411.2974 [hep-ph].

\bibitem{Panico:2011pw}
  G.~Panico and A.~Wulzer,
  JHEP {\bf 1109} (2011) 135
  [arXiv:1106.2719 [hep-ph]].
  
\bibitem{Barbieri:2007bh}
  R.~Barbieri, B.~Bellazzini, V.~S.~Rychkov and A.~Varagnolo,
  Phys.\ Rev.\ D {\bf 76} (2007) 115008
  [arXiv:0706.0432 [hep-ph]].
  
\bibitem{Grojean:2013qca}
  C.~Grojean, O.~Matsedonskyi and G.~Panico,
  JHEP {\bf 1310} (2013) 160
  [arXiv:1306.4655 [hep-ph]].
  
\bibitem{Pappadopulo:2014qza}
  D.~Pappadopulo, A.~Thamm, R.~Torre and A.~Wulzer,
  JHEP {\bf 1409} (2014) 060
  [arXiv:1402.4431 [hep-ph]].
  
\bibitem{Barbieri:2004qk}
  R.~Barbieri, A.~Pomarol, R.~Rattazzi and A.~Strumia,
  Nucl.\ Phys.\ B {\bf 703} (2004) 127
  [hep-ph/0405040].
  
\bibitem{Mrazek:2011iu}
  J.~Mrazek, A.~Pomarol, R.~Rattazzi, M.~Redi, J.~Serra and A.~Wulzer,
  Nucl.\ Phys.\ B {\bf 853} (2011) 1
  [arXiv:1105.5403 [hep-ph]].
  
\bibitem{Barbieri:2005ri}
  R.~Barbieri, T.~Gregoire and L.~J.~Hall,
  hep-ph/0509242.

\bibitem{Chacko:2005vw}
  Z.~Chacko, Y.~Nomura, M.~Papucci and G.~Perez,
  JHEP {\bf 0601} (2006) 126
  [hep-ph/0510273].
  
\bibitem{Chacko:2005un}
  Z.~Chacko, H.~S.~Goh and R.~Harnik,
  JHEP {\bf 0601} (2006) 108
  [hep-ph/0512088].
  
\bibitem{Chang:2006ra}
  S.~Chang, L.~J.~Hall and N.~Weiner,
  Phys.\ Rev.\ D {\bf 75} (2007) 035009
  [hep-ph/0604076].
  
\bibitem{Batra:2008jy}
  P.~Batra and Z.~Chacko,
  Phys.\ Rev.\ D {\bf 79} (2009) 095012
  [arXiv:0811.0394 [hep-ph]].
  
\bibitem{Craig:2013fga}
  N.~Craig and K.~Howe,
  JHEP {\bf 1403} (2014) 140
  [arXiv:1312.1341 [hep-ph]].
  
\bibitem{Craig:2014aea}
  N.~Craig, S.~Knapen and P.~Longhi,
  arXiv:1410.6808 [hep-ph].
  
\bibitem{Craig:2014roa}
  N.~Craig, S.~Knapen and P.~Longhi,
  arXiv:1411.7393 [hep-ph].
  
\bibitem{Burdman:2014zta}
  G.~Burdman, Z.~Chacko, R.~Harnik, L.~de Lima and C.~B.~Verhaaren,
  arXiv:1411.3310 [hep-ph].
  
\bibitem{Kaplan:1991dc}
  D.~B.~Kaplan,
  Nucl.\ Phys.\ B {\bf 365} (1991) 259.
  
\bibitem{Contino:2006qr}
  R.~Contino, L.~Da Rold and A.~Pomarol,
  Phys.\ Rev.\ D {\bf 75} (2007) 055014
  [hep-ph/0612048].

\bibitem{Matsedonskyi:2012ym}
  O.~Matsedonskyi, G.~Panico and A.~Wulzer,
  JHEP {\bf 1301} (2013) 164
  [arXiv:1204.6333 [hep-ph]].
\bibitem{Panico:2012uw}
  G.~Panico, M.~Redi, A.~Tesi and A.~Wulzer,
  JHEP {\bf 1303} (2013) 051
  [arXiv:1210.7114 [hep-ph]].
\bibitem{DeSimone:2012fs}
  A.~De Simone, O.~Matsedonskyi, R.~Rattazzi and A.~Wulzer,
  JHEP {\bf 1304} (2013) 004
  [arXiv:1211.5663 [hep-ph]].
  
\bibitem{Craig:2015pha}
  N.~Craig, A.~Katz, M.~Strassler and R.~Sundrum,
  arXiv:1501.05310 [hep-ph].
  
  \bibitem{Tesi}
   M.~Low, A.~Tesi and L.~T.~Wang,
  arXiv:1501.07890 [hep-ph].
  
\end{thebibliography}

\end{document}